\documentclass[12pt]{article}
\usepackage{natbib}
\usepackage{amsmath,amsfonts}
\usepackage{times} 
\usepackage{mathbbol}
\begin{document}

\title{Time, Finite Statistics, and Bell's Fifth Position}

\author{Richard D. Gill\thanks{Mathematical Institute, University
Utrecht}\thanks{gill@math.uu.nl; http://www.math.uu.nl/people/gill}}

\date{\today}

\maketitle

\abstract{In this contribution to the 2002 Vaxjo conference on the
foundations of quantum mechanics and probability, I discuss three issues connected 
to Bell's theorem and Bell-CHSH-type experiments: time and the memory loophole,
finite statistics (how wide are the error bars, under local realism?), 
and the question of whether a  loophole-free experiment is
feasible, a surprising omission on Bell's list of four positions to hold in the light
his results. L\'evy's (1935) theory of martingales, and Fisher's (1935) theory of 
randomization in experimental design, take care of time and of finite statistics.
I exploit a (classical) computer network metaphor for
local realism to argue that Bell's conclusions are independent of how one
likes to interpret probability. I give a critique of some recent anti-Bellist 
literature.}

\section*{{1. Introduction}}
It has always amazed me that anyone could find fault with 
\citet{bell64}).
Quantum mechanics cannot be cast into a classical mold.
Well, isn't that delightful? Don't Bohr, von Neumann, Feynman, all
tell us this, each in their own way?
Why else are we fascinated by quantum mechanics?
Moreover Bell writes with such economy, originality, 
modesty, and last but not least, humour.

I want to make it absolutely clear that I do not think that quantum mechanics
is non-local. Bell also made it clear that his work did not \emph{prove} that.
In fact, in \citet{bell81}, the final section of the paper
on Bertlmann's famous socks (chapter 16 of \citet{bell87}),  
he gave a list of \emph{four}
quite different positions one could take, each one logically consistent with 
his mathematical results. One of them is simply \emph{not to
care}:  go with Bohr, don't look for anything behind the scenes,
for if you do you will get stuck in meaningless paradoxes, meaningless because
there no necessity for anything behind the scenes.
If, however, like Bell himself, you have a personal preference for 
imagining a \emph{realistic} world behind the scenes,
accept with Bell that it must be \emph{non-local}. You will be in excellent company:
with Bohm-Riley, with Girardi-Rimini-Weber 
(the continuous spontaneous localization model), 
and no doubt with others. 
Alternatively, accept even worse consequences---on which more, later.

However at Vaxjo the anti-Bellists seemed to form a vociferous majority, 
though each anti-Bellist position seemed to me to be at odds with each other one.
All the same, I will in this paper outline
a recent \emph{positive} development: namely, a strengthening of 
\emph{Bell's inequality}. This strengthening does not strengthen
\emph{Bell's theorem}---quantum mechanics is incompatible with 
local realism---but it does strengthen experimental evidence for the 
ultimately more interesting conclusion: laboratory reality is 
incompatible with local realism.

You may have a completely different idea in your head from mine as to what
the phrases \emph{local realism} and \emph{quantum mechanics} stand for. 
As also was made clear at Vaxjo, a million and one different interpretations
exist  for each. Moreover these interpretations depend
on interpretations of yet other basic concepts such as \emph{probability}.
However let me describe my concrete mathematical results first, and turn to the
philosophy later. After that, I will discuss some (manifestly or not) anti-Bellist
positions, in particular those of Accardi, Hess and Philipp, 't Hooft, Khrennikov,
Kracklauer, and Volovich.

I mentioned above that \citet{bell81} lists four possible positions to hold, each
one logically consistent with his mathematical results. Naturally they were not meant to
be exhaustive and exclusive, but still I am surprised that he missed
a to my mind very interesting fifth possibility: 
namely, that any experiment which quantum
mechanics itself allows us to do, of necessity contains a loophole, preventing 
one from drawing a watertight conclusion against local realism. 
Always, \emph{because of quantum mechanics},
it will be possible to come up with
a local-realistic explanation (but each time, a different one). This logical possibility
has some support from Volovich's recent findings, and moreover makes 't Hooft's
enterprise less hopeless than the other four possibilities would suggest.
(I understand that Ian Percival has earlier promoted a similar point of view).

Personally, I do not have a preference for this position either, 
but put it forward in order to urge the experimentalists
to go ahead and prove me wrong. It is a pity that the prevailing opinion,
that the loophole issue is dead since each different loophole has been 
closed in a different experiment,  is a powerful social 
disincentive against investing one's career in doing the definitive 
(loophole free) experiment.

\section*{{2. A Computer Network Metaphor}}

To me, ``local realism'' is something which I can understand. And what I can
understand are computers (idealised, hence perfect, classical computers) 
whose state at any moment is one definite state out of some 
extremely large (albeit finite) number, and whose state
changes according to definite rules at discrete time points. Computers can
be connected to one another and send one another messages. Again, this
happens at discrete time points and the messages are large but discrete.
Computers have memories and hard disks, on which can be stored huge
quantities of information. One can store data and programs on computers.
In fact what we call a program is just data for another program (and that is
just data for another program \dots\  but not ad infinitum).

Computers can simulate randomness. Alternatively one can, in advance,
generate random numbers in any way one likes and store them on the 
hard disk of one's computer.
With a large store of outcomes of fair coin tosses (or whatever for you is the
epitome of randomness) one can simulate outcomes of any random variables
or random processes with whatever probability distributions one likes, as
accurately as one likes, as many of them as one likes,
as long as one's computers (and storage facilities) are large and fast enough.
In the last section of the paper I will further discuss whether there is any real
difference between random number generation by tossing coins or by
a pseudo-random number generator on a computer.

Computers can be cloned. Conceptually, one can take a computer and
set next to it an identical copy, identical in the sense not only that the
hardware and architecture are the same but moreover that every bit of 
information in every register, memory chip, hard disk, or whatever, is
the same.

Computer connections can be cloned. Conceptually one can collect the
data coming through a network connection, and retransmit two identical
streams of the same data.

Consider a network of five computers connected linearly. I shall call
them $\mathbb A$, $\mathcal X$, $\mathcal O$, $\mathcal Y$ and
$\mathbb B$. The rather plain  ``end'' computers $\mathbb A$ and $\mathbb B$ 
are under my control, the more fancy  ``in between'' computers
$\mathcal X$, $\mathcal O$ and $\mathcal Y$ are
under the control of an anti-Bellist friend called Luigi.
My friend Luigi has come up with a local realistic theory intended
to show that Bell was wrong, it is possible to violate the Bell inequalities
in a local realistic way. I have challenged my friend to implement his theory
in some computer programs, and to be specific I have stipulated that he
should violate the \citet*{clauseretal69} version of the Bell inequalities,
as this version is the model for the famous \citet*{aspectetal82b} experiment, and a host of
recent experiments such as that of  \citet{weihsetal98}. Moreover this experimental
protocol was certified by Bell himself, for instance in the ``Bertlmann's socks''
chapter of ``Speakable and Unspeakable'', as forming the definitive test of his theory.

Another of my anti-Bellist friends, Walter, has claimed that Bell neglected the factor 
\emph{time} in his theory. Real experiments are done in one laboratory over
a lengthy time period, and during this time period, variables at different locations
can vary in a strongly correlated way---the most obvious example being real clocks!
Well, in fact it is clear from ``Bertlmann's socks''
that Bell was thinking very much of time as being a factor in classical correlation,
see his discussion of the temporal relation between the daily number of heart-attacks
in Lyons and in Paris (the weather is similar, French TV is identical, weekend or weekday 
is the same ...).  In the course of time, the state of physical systems can drift in a
systematic and perhaps correlated way. This means that the outcomes of consecutive
measurements might be correlated in time, probability distributions are not 
stationary, and statistical tests of significance are invalidated. 
Information from the past is not forgotten, but accumulates.
The phenomenon has been named ``the memory loophole''.
More insidiously, in the course of time, information can propagate from
one physical subsystem to another, making everything even worse.
(Think of French TV, reporting events in both Paris and Lyons with a short time lag.)  
In order to accomodate \emph{time} I will allow
Luigi to let his computers communicate between themselves whatever they like,
in between each separate measurement, and I will make no demands whatsover of
stationarity or independence.  I do not demand that he simulates specific measurements
on a specific state.  All I demand is that he violates a Bell-CHSH 
inequality. I suggest that he goes for the maximal $2 \sqrt 2$ deviation corresponding to
a certain state and collection of measurement settings, but the choice is up to him,
since he has total control over his computers, and these choices are out of my
control. My computers are just going to supply the results of independent 
fair coin tosses. 

The experiment can only generate a finite amount of data. How are we going to
decide whether the experiment has proved anything? How large should $N$ be
and what is a criterion we can both agree to? A physicist would say that we have
a problem of \emph{finite statistics}.

One of my pro-Bellist friends, Gregor, an experimental physicist, has claimed that his 
experiment shows a thirty standard deviations departure from local realism.
As a statistician I am concerned that his calculation of ``thirty standard deviations''
was done assuming Poisson statistics, which comes down to assuming independence
between succesive measurements, while the anti-Bellist, because of the
memory loophole, need not buy this assumption,
hence need not buy the conclusion. As a statistician I realise that I must do my probability
calculation from the point of view of the local realist (even if in my opinion this
point of view is wrong).  I must show that,  assuming a local
realist position, the probability of such an extreme deviation as is actually observed
is very small. This is not the same as showing that, assuming quantum mechanics is
true, the probability that my experiment would have given the ``wrong''  conclusion 
(i.e., a conclusion favourable to the local realist) is very small. Of course it is a comfort to know 
this in advance of doing the experiment, and retrospectively it confirms the experimenter's
skill, but to the local realist it is just irrelevant.

Now here an interesting paradox appears: a local realist theory is typically a deterministic
theory, hence does not allow one to make probability assumptions at all. 
However I think that even local realists agree that there are situations where one
can meaningfully talk probability, even if any person's stated interpretation 
of the word might appear totally different from mine. 
However he interprets the word probability, most local
realists will agree that in a well equipped laboratory we could manufacture something
pretty close to an idealised fair coin (by which I mean a coin 
together with a well-designed coin tossing apparatus).
It could be close enough, for instance,
that we would both be almost certain that in $40\,000$ tosses 
the number of heads will not exceed 
$20\,000$  by more than $1\,000$ ($10$ standard
deviations). Behind this lies a combinatorial fact: the number of binary
sequences of length $40\,000$, in which the number of $1$'s exceeds $20\,000$
by more that $1\,000$, is less than a fraction $\exp(-\frac 12 10^2)$ 
of the total number of sequences. 

So I hope my anti-Bellist friends will let me (the person in control of computers $\mathbb A$
and $\mathbb B$) either, ahead of the experiment,
store the outcomes of fair coin tosses in them, or simulate
them with a good pseudo-random number generator, and more importantly, will be
convinced when I give probability statements concerning this and only this source of
randomness in our computer experiment.

Now here are the rules of our game. We are going to simulate an idealised, perfect (no 
classical loopholes) Bell-CHSH type delayed choice experiment. For the sake of argument
let us fix $N=15\,000$ as the total number of \emph{trials} (pairs of events, photon pairs, \dots).
In advance, Luigi has set up his
three computers with any programs or data whatsoever stored on them. He is allowed to
program his chameleon effect, or Walter's B-splines and 
hidden-variables-which-are-not-actually-elements-of-reality, or Al's theory 
of QEM, whatever he likes. 

For $n=1,\dots,N=15\,000$, consecutively, the following happens:
\begin{itemize}
\item[1.] Computer $\mathcal O$, which we call the \emph{source}, sends information to
computers $\mathcal X$ and $\mathcal Y$, the \emph{measurement stations}. It can be
anything. It can be random (previously stored outcomes of actual random experiments)
or pseudo-random or deterministic. It can depend in an arbitrary way on the results of
past trials (see item 5). Without loss of generality it can be considered to
be the same---send to each computer, both its own message and the message for the other.
\item[2.] Computers $\mathbb A$ and $\mathbb B$, which we call the \emph{randomizers}, 
each send a 
\emph{measurement-setting-label}, namely a $1$ or a $2$, to computers 
$\mathcal X$ and $\mathcal Y$. Actually, I will generate the labels to simulate 
independent fair coin tosses
(I might even use the outcomes of real fair coin tosses, done secretly in advance and 
saved on my computers' hard disks).
\item[3.] Computers $\mathcal X$ and $\mathcal Y$ each output an outcome $\pm 1$, computed
in whatever way Luigi likes from the available information at each measurement station.
He has all the possibilities mentioned under item 1. What each of these two computers 
do not have, is the measurement-setting-label which was delivered to the other.
Denote  the outcomes $x^{(n)}$ and $y^{(n)}$.
\item[4.] Computers $\mathbb A$ and $\mathbb B$ each output the measurement-setting-label
which they had previously sent to $\mathcal X$ and $\mathcal Y$. Denote these labels
$a^{(n)}$ and $b^{(n)}$. An independent referee will confirm that these are identical to
the labels given to Luigi in item 2.
\item[5.] Computers $\mathcal X$, $\mathcal O$ and $\mathcal Y$ may communicate with
one another in any way they like. In particular, all past setting labels are available at all
locations. As far as I am concerned, Luigi may even alter the 
computer programs or memories of his machines. 
\end{itemize}

At the close of these $N=15\,000$ trials we have collected $N$ quadruples
$(a^{(n)},b^{(n)},x^{(n)},y^{(n)})$, where the measurement-setting-labels take the values
$1$ and $2$, the measurement outcomes take the values $\pm 1$.
We count the the number of times the two outcomes were equal to one another, and
the number of times they were unequal, separately for each of the four possible
combinations of measurement-setting-labels:
\begin{align*}
N_{ab}^=~&=~\#\{n:x^{(n)}=y^{(n)},\,(a^{(n)},b^{(n)})=(a,b)\},\\
N_{ab}^{\ne} ~&=~\#\{n:x^{(n)}{\ne} y^{(n)},\,(a^{(n)},b^{(n)})=(a,b) \},\\
N_{ab}~&=~\#\{n:(a^{(n)},b^{(n)})=(a,b)\}.
\end{align*}
From these counts we compute
four empirical \emph{correlations} (a mathematical statistician would call them 
raw, or uncentred, product moments), as follows.
\begin{equation*}
\widehat \rho_{ab}~=~\frac{N_{ab}^=-N_{ab}^{\ne} }{N_{ab}}.
\end{equation*}
Finally we compute the CHSH contrast
\begin{equation*}
\widehat S~=~\widehat\rho_{12}-\widehat\rho_{11}-\widehat\rho_{21}-\widehat\rho_{22}.
\end{equation*}
Luigi's aim is that this number is close to $2\sqrt 2$, or at least, much larger than $2$.
My claim is that it cannot be much larger than $2$; in fact, I would not expect
a deviation larger than several times $1/\sqrt N$ above $2$. \citet{weihsetal98}
obtained a value of $\widehat S\approx 2.73$ also with $N\approx 15\,000$ in an experiment
with a similar layout, except that the measurement stations were polarizing beam-splitters
measuring pairs of entangled photons transmitted from a source through $200\text{m}$ of
glass fibre each, and the randomizers were quantum optical devices simulating (close to)
fair coin tosses by polarization measurements of completely unpolarized photons,
see Appendix 1.
A standard statistical computation showed that the value of $\widehat S$ they found is
$30$ standard deviations larger than $2$.

Please note that Luigi's aim is certainly achievable from a logical point of view.
It is conceivable, even,  that $N_{12}^{\ne} =0$ and $N_{11}^= =N_{21}^= =N_{22}^= =0$, 
hence that
$\widehat\rho_{12}=+1$, $\widehat\rho_{11}=\widehat\rho_{21}=\widehat\rho_{22}=-1$,
and hence that $\widehat S=4$. In fact if Luigi would generate his outcomes just as I generated
the settings, as independent fair coin tosses, this very extreme result does have a 
positive probability. The reader might like to compute the chance. 

In order to be able to make a clean probability statement, I would like to 
make some harmless modifications to $\widehat S$. First of all, note that the ``correlation''
between binary ($\pm1$ valued) random variables
is twice the probability that they are equal, minus $1$:
\begin{equation*}
\widehat \rho_{ab}~=~\frac{N_{ab}^=-(N_{ab}-N_{ab}^=)}{N_{ab}}~=~2\frac{N_{ab}^=}{N_{ab}}-1.
\end{equation*}
Define 
\begin{equation*}
\widehat p_{ab}^= ~=~N_{ab}^=/N_{ab}.
\end{equation*}
Luigi's aim is to have
\begin{equation*}
(\widehat S-2)/2~=~\widehat p_{12}^=-\widehat p_{11}^=-\widehat p_{21}^=-\widehat p_{22}^=
\end{equation*}
close to
$\sqrt 2 -1$, my claim is that it won't be much larger than $0$. Now multiply $(\widehat S-2)/2$
by $N/4$ and note that the four denominators $N_{ab}$ in the formulas for the 
$\widehat p_{ab}^=$
will all be pretty close to the same value, $N/4$. 
I propose to focus on the quantity $Z\approx N(\widehat S -2)/8$ obtained
by cancelling the four denominators against $N/4$:
\begin{equation*}
Z~=~N_{12}^= -N_{11}^= -N_{21}^= -N_{22}^= .
\end{equation*}
Luigi's aim is to have this quantity close to $N(\sqrt 2 -1)/4\approx N/10$, or at least,
significantly larger than $0$, 
while I do not expect it to be larger than $0$ by several multiples of $\sqrt N$.

He will not succeed. It is a theorem that
\emph{whatever Luigi's programs and stored data}, and \emph{whatever
communication between them at intermediate steps},
\begin{equation*}
\Pr\Bigl\{Z\ge k\sqrt N\Bigr\}~\le~\exp\Bigl(-{\textstyle \frac 12} k^2\Bigr ),
\end{equation*}
where $k\ge 0$ is arbitrary. For instance, with $N=15\,000$, and $k=12.25$,
one finds that $k\sqrt N\approx N/10$ while $\exp(-\frac 12 12.25^2) \le 10^{-32}$.

In fact I can improve this result---as if improvement were necessary!---replacing $k$ 
in the \emph{right hand side} by  a number one and three quarters times as large, 
by a technique 
called random time change, which I shall explain later.
But I \emph{cannot} get any further improvement, in particular, I cannot reach
$\exp(-\frac 12 30^2)$, corresponding to \citeauthor{weihsetal98}'s
(\citeyear{weihsetal98}) thirty standard 
deviations. Why? Because their calculation (with $N\approx 15\,000$)
was done assuming independent and identically distributed trials, and assuming
probabilities equal to the observed relative frequencies, very close to those predicted
by quantum mechanics; whereas my calculation is done \emph{assuming local realism},
under the most favourable conditions possible under local realism, and assuming no
further randomness than the independent fair coin tosses of the randomizers.

If you are unhappy about my move from correlations to counts, let me just
say that I can make similar statements about the original $\widehat S$, by 
combining the probability inequality for $Z$ with similar but easier probability
inequalities for the $N_{ab}$.

\section*{{3. Martingales}}

Let me give a sketch of the proof. 
I capitalize the symbols for the settings and outcomes because I am thinking of them
as random variables.
Write each of the counts in the expression for $Z$
as a sum over the $N$ trials of an indicator variable (a zero/one valued random variable)
indicating whether or not the event to be counted occured on the $n$th trial. 
A difference
of sums is a sum of differences. Consequently, if we define
\begin{align*}
\Delta_{ab}^{(n)}~&=~\mathbb 1\{X^{(n)}=Y^{(n)},\,(A^{(n)},B^{(n)})=(a,b)\} ,\\
\Delta^{(n)}~&=~\Delta_{12}^{(n)}-\Delta_{11}^{(n)}-\Delta_{21}^{(n)}-\Delta_{22}^{(n)},\\
Z^{(n)}~&=~\sum_{m=1}^n \Delta^{(m)},
\end{align*}
then $Z=Z^{(N)}$. Now I will show in a moment, 
using a variant of  \citeauthor{bell64}'s \citeyear{bell64} argument,
that for each $n$, conditional on the past history of the first $n-1$ trials, 
the expected value of $\Delta^{(n)}$
does not exceed $0$, whatever that history might be. 
Moreover, $\Delta^{(n)}$ can only take on the values $-1$, $0$
and $1$, in particular, its maximum minus its minimum possible value (its range)
is less than or equal to $2$.
This makes the stochastic process $Z^{(0)},Z^{(1)},\dots,Z^{(n)},\dots,Z^{(N)}$ a
\emph{supermartingale} with \emph{increments in a bounded range}, 
and with initial value $Z^{(0)}=0$.
The definition of a supermartingale
is precisely the property that the increments $\Delta^{(n)}$ have nonpositive conditional
expectation given the past, for each $n$. A supermartingale is a generalisation of a random
walk with zero or negative drift. Think for instance of the amount of money in your pocket as you
play successive turns at a roulette table, where the roulette wheel is perfect, but the presence
of a \texttt{0} and \texttt{00} means that on average, whatever amount you stake, and whatever you
bet on, you lose $1/19$ of your stake at each turn. You may be using some complex
or even randomized strategy whereby the amount of your stake, 
and what you bet on (a specific number, or
red versus black, or whatever) depends on your past experience and on auxiliary random
inputs but still you lose
on average, conditional on the past at each time point, whatever the past. 
The capital of the bank is a submartingale---nonnegative drift. 
If there would be no \texttt{0} and \texttt{00},
both capitals would be martingales---zero drift.
In a real roulette game there will be a maximum stake and hence a maximum payoff.
Your capital changes by an amount between the maximal payoff and minus 
the maximal stake. Thus your capital while playing roulette develops in
time as a supermartingale with increments of bounded range (maximal payoff plus
maximal stake).
If you cannot play more than $N$ turns,
with whatever strategy you like (including stopping early),
it can only very rarely happen that your capital increases by more than a 
few times $\sqrt N$ times half the range, as we shall now see.

According to \citeauthor{hoeffding63}'s \citeyear{hoeffding63} inequality,
if a supermartingale $(Z^{(n)}:n=0,1,\dots,N)$ is zero at time $n=0$,
and the range of its increments is bounded by $2$, then
\begin{equation*}
\Pr\Bigl\{\max_{n\le N} Z^{(n)} \ge k\sqrt N\Bigr\}~\le~\exp\Bigl(-{\textstyle \frac 12} k^2\Bigr ).
\end{equation*}
Note that if the increments of the supermartingale were actually independent and
identically distributed, with range bounded by $2$, then the maximum variance of 
$Z^{(N)}$ is precisely equal to $N$, achieved when the increments are equal to $\pm 1$
with equal probability $\frac 12$. The Chebyshev inequality 
(sometimes known as Markov inequality) would then tell us that $Z^{(N)}$ exceeds 
$k\sqrt N$ with probability smaller than $1/k^2$. Hoeffding has improved this in two
ways: an exponentially instead of geometrically decreasing probability, and a maximal
inequality instead of a pointwise inequality. 
One cannot do much better than this inequality: in the
most favourable case, just described, for large $N$ we would have that $Z^{(N)}$
is approximately normally distributed with variance $N$, and the probability of
large deviations of a normal variate behaves up to a constant and a lower order 
(logarithmic) term  precisely like $\exp(-\frac 12 k^2)$.

The proof of Hoeff\-ding's inequality can be
found in the better elementary probability textbooks and uses Markov's inequality,
together with a random time change argument, and finally 
some elementary calculus.
This gives a clue to how I can improve the result: consider the random process
only at the times when $\Delta^{(n)}\ne 0$. In other words, thin out the time points
$n=0,1,...$ in a random way, only look at the process at the time points which are left.
By Doob's optional stopping theorem it is \emph{still} a supermartingale 
when only looked at intermittently, even when we only
look at random time points, provided that we never need to look ahead to select these
time points.
The increments of the thinned process still have a range bounded by $2$.
Hence Hoeffding's inequality still applies. However, time is now running faster,
thus the value of $N$ in the inequality as stated for the new process corresponds to
$cN$ in the old, with $c>1$. In fact, in the actual experiment we only see a $\pm 1$
in a fraction $0.325=\frac 14 \sum_{a,b}p^=_{a,b}$ of all trials, 
hence we can improve the $k$ on the right hand
side by a factor $1/\sqrt 0.325=1.75$, hence $12.25$ can be increased to $21.5$.

It remains to prove the supermartingale property. Consider the quantity $\Delta^{(n)}$.
Condition on everything which happened in the first $n-1$ trials, and also on 
whatever new information Luigi placed on his computers between the $n-1$st and $n$th
trial.
Consider the situation just after Luigi's computers $\mathcal X$ and $\mathcal Y$ have
received their information from $\mathcal O$, just before they receive the settings
from $\mathbb A$ and $\mathbb B$.  Under my conditioning, the
state of Luigi's computers is fixed (non random). Clone Luigi's computers
(this is only a thought experiment). Give the first copy
of computer $\mathcal X$ the input $1$, as if this value came from $\mathbb A$,
and give the second copy the input $2$; 
do the same in the other wing of the experiment. Let's drop the upper
index ${(n)}$, and denote by $x_1$ and $x_2$ the outputs of the two clones of $\mathcal X$,
denote by $y_1$ and $y_2$ the outputs of the two clones of $\mathcal Y$.
Because we are conditioning on the past up till the generation of the settings in the $n$th
trial, everything is deterministic except the two random setting labels,
denoted by $A$ and $B$.
The actual output  from the actual (uncloned) computer $\mathcal X$ is $X=x_A$, 
similarly in the other wing of the experiment. We find
\begin{align*}
\Delta^{(n)}_{ab}~&=~\mathbb 1\{X=Y,\,(A,B)=(a,b)\}\\
~&=~\mathbb 1\{x_a=y_b\}\mathbb 1\{(A,B)=(a,b)\}.
\end{align*}
The (conditional) expectation of this quantity is $\mathbb 1\{x_a=y_b\}/4$, since the
randomizers still produce independent fair coin tosses given the past, and
given whatever further modifications Luigi has made.
Hence the expectation of $\Delta^{(n)}$
given the past up to the start of the $n$th trial equals one quarter times
$\mathbb 1\{x_1=y_2\}-\mathbb 1\{x_1=y_1\}-\mathbb 1\{x_2=y_1\}-\mathbb 1\{x_2=y_2\}$.
Now since the $x_a$ and $y_b$ only take the values $\pm 1$, it follows that
$(x_1y_2)(x_1y_1)(x_2y_1)(x_2y_2)=+1$. The value of a product of two $\pm1$ valued
variables encodes their equality or inequality. We see that the number of equalities
within the four pairs involved is even. It is not difficult to see that it follows from this,
that the value of
$\mathbb 1\{x_1=y_2\}-\mathbb 1\{x_1=y_1\}-\mathbb 1\{x_2=y_1\}-\mathbb 1\{x_2=y_2\}$
can only be $0$ or $-2$, so is always less than or equal to $0$. We have proved the required 
property of the conditional expectation of $\Delta^{(n)}$ conditioning not only on the
past $n-1$ trials but also on what happens between $n-1$st and $n$th trial. 
Average over all possible inter-trial happenings, to obtain the result we want. 
The theorem is proved.

As I remarked before, computers $\mathcal X$, $\mathcal O$ and $\mathcal Y$ are
allowed to communicate in anyway they like between trials, and Luigi is even allowed
to intervene between trials, changing their programs or data as he likes, even 
in a random way if he likes. He can make use in all his computers of the outcomes
of the randomizers at all previous trials. It does not help. No assumption has been made
of any kind of long run stability of the outcomes of his computers, or stationarity of
probability distributions. The only requirement has been on my side, that I am allowed to
choose setting labels at random, again and again. Only this randomness drives my conclusion.
You may see my theorem as a  combinatorial statement, referring to the fraction of results
obtained under all the $4^N$ different combinations of values of all $a^{(n)}$ and $b^{(n)}$.

Further details are given in \citet{gill03} though there I used the Bernstein rather than the
Hoeffding inequality; Hoeffding turned out to give sharper results. A publication is
in preparation giving more mathematical details and further results.
In particular one can give similar Hoeffding bounds for the original quantity of interest
$\widehat S$, and the unbiasedness of the two randomizers is not crucial. In
fact Weihs had probabilities of heads equal to $0.48$ and to $0.42$ in the two
wings of his experiment.

Martingales (avant la lettre) were 
introduced into probability theory by the great French probabilist Paul L\'evy in 1935.
The name martingale was given to them a few years later by his student Ville, 
who used them to effectively destroy Richard von Mises' programme to found probability on the
notion of collectives and limiting relative frequencies. Only Andrei Nikolaevich Kolmogorov
realized that this conclusion was false, and he went on to develop the notion of 
computational complexity based on von Mises' ideas. Later still, the Dutch mathematician
Michiel van Lambalgen has shown that a totally rigorous mathematical 
theory of collectives can be derived if one replaces the axiom of choice (which makes
mathematical existence theorems easy, a double edged sword since it creates pathologies
as well as desired results) with an alternative axiom, closer to physical intuition.

The year 1935 also saw the introduction, by the great British statistician 
Sir Ronald Aylmer Fisher, of the notion of randomization into experimental design.
He showed that randomized designs gave
an experimenter total control of uncontrollable factors which could otherwise
prevent any conclusions being drawn from an experiment.

\section*{{4. Metaphysics}}

The interpretation of Bell's theorem depends on
notions of what is quantum mechanics, what is local realism, and behind them, what
is probability. By the way, Bell himself does not state a theorem; just shows that 
certain assumptions imply a certain inequality. He shows that under a conventional
interpretation of quantum mechanics, this inequality could be violated. However,
it has become conventional to call the statement that quantum mechanics and local
realism are incompatible with one another, Bell's theorem. This is a very convenient
label,  all the more convenient since later authors have obtained the same conclusion 
through consideration of other predictions of quantum mechanics, some
of them not on the face of it involving an inequality as Bell's. 
Actually, \citet*{gilletal03} argue elsewhere that these proofs of
Bell's theorem without inequalities \citep{hardy93}, or even without probability
\citep*{greenbergeretal89}, do actually involve
hidden probability inequalities.

On the one hand, Bell's theorem depends on an interpretation of quantum mechanics,
together with an assumption that certain states and measurements, which one can
consider as allowed by the mathematical framework, can also arise ``in Nature'',
including Nature as manipulated by an experimenter in a laboratory.
What I call Bell's missing fifth position, is the position that quantum mechanics
itself forbids these states ever to exist. And not just the specific states and measurements
corresponding to a particular proof of Bell's theorem, but any which one could use in the
proof. Restricting attention to a Bell-CHSH type experimental set-up,
one does not need to achieve the
magic $2\sqrt 2$, one only needs to significantly exceed the bound $2$.
However, let me briefly describe the calculations behind this magic number (an 
upper bound under quantum mechanics, according to the Cirel'son inequality),
since this leads naturally to a discussion of the role of probability.

It is conventional and reasonable to take the Hilbert space corresponding to a physical
system consisting of two well separated parts of space as being the tensor product of
spaces corresponding to the two parts separately. To achieve $2\sqrt 2$,
we need that a state exists
(can be made to exist)
of the joint system, which can be written (approximately)
in the form $|00\rangle+|11\rangle$
(up to normalization, and up to a tensor product with whatever else you like, pure or mixed);
where as usual $|00\rangle=|0\rangle\otimes|0\rangle$, 
$|11\rangle=|1\rangle\otimes|1\rangle$, and $|0\rangle$ and $1\rangle$ stand for
two orthonormal vectors in both the first and the second space.
We need that one can simultaneously (to a good enough approximation)
measure whether the first subsystem is in
the state $\cos\alpha|0\rangle+\sin\alpha|1\rangle$ or in the state
orthogonal to this, $\sin\alpha|0\rangle-\cos\alpha|1\rangle$; and whether the second
is in the state $\cos\beta|0\rangle+\sin\beta|1\rangle$ or in 
$\sin\beta|0\rangle-\cos\beta|1\rangle$; where one may choose between $\alpha=\alpha_1$
or $\alpha=\alpha_2$, and between $\beta=\beta_1$ and $\beta=\beta_2$; and
where a good choice of angles (settings) leading to the famous $2\sqrt 2$ are
$\alpha_1=-\pi/4-\pi/8$, $\alpha_2=-\pi/8$, $\beta_1=0$, $\beta_2=\pi/4$.

Conventionally it is agreed that  the probability to find subsystem one in state $|\alpha\rangle=\cos\alpha|0\rangle+\sin\alpha|1\rangle$  and
subsystem two in state $|\beta\rangle=\cos\beta|0\rangle+\sin\beta|1\rangle$, when prepared
in $\Psi=(|00\rangle+|11\rangle)/\sqrt 2$, is the squared length of the inner product
of $\Psi$ with $|\alpha\rangle\otimes|\beta\rangle$, which turns out to equal
$\frac12\cos^2(\alpha-\beta)$. This is the probability of the outcome $+1,+1$. The probability
of $-1,-1$ turns out to be the same, while that of $+1,-1$ and of $-1,+1$ are both equal
to $\frac12\sin^2(\alpha-\beta)/2$. The marginal probabilities of $\pm 1$ now turn out to
equal $\frac12$ and the probability of equal outcomes is $\cos^2(\alpha-\beta)$.
Under the choices of angles above, one obtains $p_{12}^= =(1+1/\sqrt 2)/2\approx 0.85$,
while all the other $p_{ab}^= =(1-1/\sqrt 2)/2\approx 0.15$. Consequently
$p_{12}^=-p_{11}^=-p_{21}^=-p_{22}^= = (\sqrt 2 -1)$.

Out of these calculations came \emph{joint probabilities of outcomes of binary 
measurements} and every word here needs to be taken literally, if the argument is
to proceed: there are measurements taken in both wings of the experiment, and each
can only result in a $\pm 1$. We use quantum mechanics to tell us what the probability
of various combinations of outcomes is. Now there are a great many ways to try
to make sense of the notion of probability, but everyone who uses the word in the
context of quantum mechanics would agree that if one repeatedly measures a quantum
system in the same state, in the same way, then relative frequencies of the various
possible outcomes will stabilize in the long run, and they will stabilize to
the probabilities, whatever that word may mean, computed by quantum mechanics.
In the quantum version of our experiment, 
$Z/N$ will stabilize to the value $(\sqrt 2 -1)/4$.

My mathematical derivation of a stronger (probabilistic) version of the Bell inequalities
did not hinge on any particular interpretation of probability. Someone who uses the word
probability has a notion of fair coin tosses, and will not hesitate to apply probability
theory to experiments involving nothing else than two times $15\,000$ fair
coin tosses. If a certain event specified before the coins are tossed has a probability
smaller than $10^{-32}$ one is not going to see that event happen (even though
logically it \emph{might} happen).

It seems to me that the \emph{interpretation} of probability does not play any serious role
in the ongoing controversy concerning Bell's theorem. What does play a role is that 
quantum mechanics is used to compute joint probabilities of outcomes
of binary measurements.  

Many quantum physicists will object that real physicists do not use
quantum mechanics to compute probabilities, only the certain values of averages 
pertaining to huge collectives. Many others avoid recourse to Born's law by extending 
the quantum mechanical
treatment to as large a part of the measurement device as possible. If probability is 
involved it appears to come in through an uncontroversial backdoor as statistical variation 
in the medium or the elements of the collective.

That may be the situation in many fields, but people in those fields do not then 
test Bell's theorem. The critical experiment involves binary outcomes and 
binary settings, committed to sequentially as I have outlined.
A better objection is that in no experiment done to date,
has the experimental protocal described in my computer metaphor been literally
enforced. For instance, in \citeauthor{weihsetal98}'s
(\citeyear{weihsetal98}) experiment, the only one to date where
the randomization of detector settings at a sufficiently fast rate 
was taken seriously (Aspect did his best but could only implement a poor surrogate), 
the $N=15\,000$ events were post-selected from an enormously much larger 
collection of small time intervals in most of which there was no detection event at 
all  in either wing of the experiment; in a small proportion there was one detection event in one
wing of the experiment or the other but not both; and in a smaller proportion still,  
there was a detection
event in both wings of the experiment. Bell's argument just does not work when the
binary outcomes are derived from a post-experimental conditioning (post-selection)
on values of other variables.  Other experiments free of this loophole, did not
(and could not) implement the delayed random choice of settings; for instance
\citeauthor{roweetal01}'s
(\citeyear{roweetal01}) experiments with trapped ions.

Bell was well aware of this problem. In ``Bertlman's socks''  he offers a re\-solution, whereby
the source $\mathcal O$ may output at random time moments a signal that something is
about to happen. Measurements at $\mathcal X$ and $\mathcal Y$ based on a stream
of random settings from $\mathbb A$ and $\mathbb B$ take place continuously, but
after the experiment has run for some time, one selects just those measurements within
an appropriate time interval after a saved ``alert'' message from $\mathcal O$. It is
practically extremely important that this selection may be done \emph{after} the 
experiment has run its course. Post-selection is bad, but post-pre-selection is fine.

By the way, the martingale methods I outlined above are admirably suited to adaptation and
extension to continuous time measurement (of discrete events). Under reasonable 
(but of course untestable) ``unbiased detection''  assumptions, 
one can obtain the same kind of inequalities, but now allowing detection
events at random time points, and a random total number of events.

But is  ``local realism''  adequately represented by
my metaphor of a computer network?
For Bell, the key property of the crucial experiment is
that the measurement station $\mathcal X$ commits itself to a specific (binary) outcome, 
shortly after receiving a (binary) input from randomizer $\mathbb A$, before a signal from
the other wing of the experiment could have arrived with information concerning the
input which randomizer $\mathbb B$ generated in the other wing of the experiment.
In the short time period between input of $a$ and output of $x$, as far as the physical
mechanism leading to the result $x$ is concerned, we need only consider a bounded region of
space which completely excludes the physical systems $\mathbb B$ and $\mathcal Y$.
For me, ``local realism'' should certainly imply  that a sufficiently
detailed (microscopic) specification of the state in some bounded region of space 
would (mostly) fix the outcomes of  macroscopic, discrete (for instance, 
binary) variables.  For instance, a sufficiently detailed specification of the initial 
state of a coin-tossing apparatus would (mostly) fix the outcome. This does not prevent
the outcome from being apparently random, on the contrary, but it does ``explain'' the
apparent randomness through the variation of the initial conditions when the experiment
is repeated.

This means that in a thought experiment one can clone the relevant aspects of the
relevant portion of physical space, and one can carry out the thought experiment: feed
into the same physical system both of the possible inputs from the randomizer and 
thereby fix both the possible outputs. The output you actually see is what you would have
seen if you would have chosen, the input which you actually chose.

\citet{bell64,bell87} used a statistical conditional independence assumption,
together with an assumption that conditional probability distributions of outcomes
in one wing of the experiment do not depend on settings in the other wing, 
rather than my ``counterfactual definite''
characterization of local realism. Actually it is a mathematical theorem that the
two mathematical notions are equivalent to one another. Each implies the other.
Note that I do not require that my counterfactual or hidden variables physically
exist, whatever that might mean, but only that they can be mathematically 
introduced in such a way that the mathematical model with ``counterfactuals''
reproduces the joint probability distribution of the manifest variables.

In my opinion the present unfashionableness of counterfactual reasoning in
the philosophy of science is quite misguided. We would not have ethics,
justice, or science, without it.

The original EPR argument also gives support for these counterfactuals:
we know that if one measures with the same settings in the two wings of the
experiment, one would obtain the same outcomes. Hence a local realist
(like Einstein) quite reasonably 
considers the outcome which one would find under a given setting in one wing of
the experiment,
as deterministically encoded in the physical state of that part of the physical system, 
just before it is measured, independently of how it is actually measured.

In my opinion the stylized computer network metaphor for a good Bell-CHSH type
experiment is precisely what Bell himself was getting at. One cannot attack Bell on
the grounds that this experiment has never been done yet. One might attack him
on the grounds that it never can be done. One will need good reasons for this.
His argument does not require photons, nor this particular state and these particular
measurements. Again, showing that a particular experimental set-up 
using a particular kind of physical system is unfeasible,
does not show that all experimental set-ups are unfeasible.

\section*{{5. A Miscellany of Anti-Bellist Views}}

\subsection*{{Bell's Four Positions}}

Bell offered four quite different positions which one might like to take 
compatible with his mathematical results. They were:
\begin{itemize}
\item[1.] Quantum mechanics is wrong.
\item[2.] Predetermination.
\item[3.] Nature is non-local.
\item[4.] Don't care (Bohr)
.
\end{itemize}
In my opinion he missed an intriging fifth position:
\begin{itemize}
\item[5.] A decisive experiment \emph{cannot} be done.
\end{itemize}
I would like to discuss a  number of recent works in the light of these possibilities
and the results I have described above.

\subsection*{{Accardi and the Chameleon Effect}}

In numerous works L.~Accardi claims that Bell's arguments are fundamentally flawed,
because Bell could only think of randomness in a classical way: pulling coloured
balls out of urns, where the colour you get to see was the colour which was already
painted on the ball you happended to pick. If however you select a chameleon out
of a cage, where some chameleons are mutant, and you place the chameleon on a
leaf, it might turn green, or it might turn brown, but it certainly did not have that colour
in advance. 

This is certainly a colourful metaphor but I do not think that chameleons are that different
from coloured billiard balls: according to Accardi's own story whether or not a chameleon
is mutant is determined by its genes, which certainly did not get changed by picking up
one chameleon or another; and a mutant chameleon always turns brown when placed
on a green leaf.

The metaphor is also supposed to carry the idea that the measurement outcome is not
a preexisting property of the object, but is a result of an evolution of measurement
apparatus and measured object together. It seems to me that this is precisely Bohr's
Copenhagen interpretation: one cannot see measurement outcomes separate from
the total physical context in which they appear. Bohr's answer to EPR was to apply
this idea also rigorously even when two parts of the measurement apparatus and
two parts of the object being measured are light-years apart. This philosophy
certainly abolishes the EPR paradox but to my mind hardly explains it.

Accardi does provide some mathematics (of the quantum probability kind)
which is supposed to provide a local realistic model of the EPR phenomenon.
Naturally a good quantum theoretician is able to replace the von Neumann measurement
of one photon by a Schr\"odinger evolution of a photon in interaction with a
measurement device in such a way that though particle and apparatus together
are still in a pure state at the end of the evolution, the reduced state of the
measurement apparatus is a mixed state over two macroscopically distinct
possibilities. One can do this for the two particles simultaneously and arrive at
a mathematical model which reproduces the EPR correlations in a local way,
in a sense that the various items in the model can be ascribed to separate parts
of reality.  I don't think it qualifies as a local realistic model.

However Accardi believes it is a local realistic model in the sense that he could
have computer programs running on a network of computers which would
simulate the EPR correlations, while implementing his mathematical theory.
These computer programs have run through several versions but presently
Accardi's web site does not seem to be accessible.  Unfortunately none of
the versions I have been able to test allowed me the sort of control over the
protocol of the experiment, to which I am entitled under. 
In particular, I was not able to
see the raw data, only correlations. However by setting $N=1$ one can get
some idea what is going on inside the blackbox. Surprisingly with $N=1$
it was possible to observe a correlation of $\pm1.4$. Has the chameleon  multiplied
the outcome $\pm 1$ in one wing of the experiment by $\sqrt 2$?
A later version of the program also allowed the
outcome ``no detection'' and though the author still claims categorically that
Bell was wrong, the main thrust of the paper seems now to be to model actual
experiments, which as is abundantly known suffer seriously from the detection
loophole.

The martingale results which I have outlined above were derived in order to
determine how large $N$ should be, so that I would have no danger of
losing a public bet with Accardi, that his computer programs could not
violate the Bell-CHSH inequalities in an Aspect-type experiment, which is
to say an experiment with repeated random choice of settings. Since he was
to be totally free in what he put on his computers I could not use standard
statistical methods to determine a safe sample size. Fortunately the 
martingale came to my rescue.

\subsection*{{Hess and Philipp and non-elements-of-reality}}

I first became aware of the contributions of Hess and Philipp through an article
in the science supplement of a reliable Dutch newspaper. Einstein was right after
all. It intrigued me to discover that there was a fatal time-loophole in Bell's
theorem, when I had just succeeded in fixing this loophole myself in order to
make a safe bet with Accardi. 

The first publications by these authors appeared in print in somewhat mangled form,
since the journal had requested that the paper be reduced and cut in two pieces.
Some notational confusions and mismatches made it very difficult to follow the 
arguments. On the one hand the papers contained a long verbal critique of Bell,
on the lines that correlations at a distance can easily be caused by synchronous
systematic variation in other factors. This is Bell's own story of the frequency of
heart attacks in distant French cities. On the other hand the papers contained a
highly complex mathematical model which was supposed to represent a local
realistic reproduction of the singlet correlations. Unfortunately the authors
chose only to verify some necessary conditions for the locality of their model.
Hidden variables which in the model were supposed to ``belong'' to one 
measurement station or the other were shown to be statistically independent 
of one another.

In the latest publication Hess and Philipp have given a more transparent
specification of their model, and in particular have recognised the important role
played by one variable which in their earlier work was either treated as a mere
index or even suppressed from the notation altogether. This variable is supposed
to represent some kind of micro-time variable which is resident in both wings
of the experiment. It turns out to have a probability distribution which depends
on the measurement settings in both wings of the experiment. The authors
implicitly recognise that it is non-local but christen it a ``non-element-of-reality''.
Thus non-local hidden variables are fine, we just should not think of them as
being \emph{real}.  They wisely point out that it seems to be a very difficult 
problem to decide which variables are elements or reality and which are not.
In Appendix 2, I give a simplified version of their model.

\subsection*{{'t Hooft and predetermination}}

't Hooft notes that at the Planck scale experimenters will not
have much freedom to choose settings on a measurement apparatus.
Thus Bell's position 2 gives license to search for a classical, local,
deterministic theory behind the quantum mechanical theory of the world
at that level. So far so good. 

However, presumably the quantum mechanical theory of the world at the Planck
scale is the foundation from which one can derive the quantum
mechanical theory of the world at levels closer to our everyday experience.
Thus, his classical, local and deterministic theory for physics at the Planck scale
is a classical, local and deterministic theory for physics at the level of
present day laboratory experiments testing Bell's theorem. It seems to me that
there are now two positions to take. The first one is that there is, also at our
level, no free choice. The experimenter thinks he is freely choosing setting label
number 2 in Alice's wing of the experimenter, but actually the photons arriving
simultaneously in the other wing of the experiment, or the stuff of the measurement
apparatus there, ``know''  this in advance and capitalize on it in a very clever way:
they produce deviations from the Bell inequality, though not larger than
Cirel'sons quantum bound of $2\sqrt 2$ (they are, after all, bound by 
quantum mechanics).  But we have no way of seeing that our ``random'' 
coin tosses are not random at all, but are powerfully correlated with forever hidden variables 
in measurement apparatus far away.  I find it inconceivable that there is such 
powerful coordination between such totally different physical systems (the brain of
the experimenter, the electrons in the photodetector, the choice of a particular
number as seed of a pseudo-random number generator in a particular
computer program) that Bell's inequality can be resoundingly violated in the
quantum optics laboratory, but nature as a whole appears ``local'', and 
randomizers appear random.

Now ``free choice''  is a notion belonging to philosophy and I would prefer not to
argue about physics by invoking a physicist's apparently free choice. It is a fact
that one can create in a laboratory something which looks very like randomness.
One can run totally automated Bell-type experiments in which measurement
settings are determined by results of a chain of separate physical systems
(quantum optics, mechanical coin tossing, computer pseudo-random number
generators). The point is that if we could carry out a perfect and succesful Bell-type
experiment, then if local realism is true 
an exquisite coordination persists throughout this complex of physical systems
delivering precisely the right measurement settings at the two locations to
violate Bell's inequalities, while hidden from us in all other ways.

There is another position, position 5: the perfect Bell-type experiment cannot
be made. Precisely because there is a local realistic hidden layer to the deepest
layer of quantum mechanics, when we separate quantum-entangled physical 
systems far enough from one another in order to do separate and randomly
chosen measurements on each, the entanglement will have decayed so far
that the observed correlations have a classical explanation. Loopholes
are unavoidable and the singlet state is an illusion.

\subsection*{{Khrennikov and exotic probability theories}}

In a number of publications Khrennikov constrasts a classical probability
view which he associates with Kolmogorov, with a so-called
contextualist viewpoint.  He also contrasts  the Kolmogorov point
of view and the von Mises (frequentist). Furthermore, he has suggested that 
the resolution of Bell's paradox might be found in some non-standard probability
theory, for instance $p$-adic. A rationale for this might be that
stabilization of relative frequencies might not be a fact at the micro-level, 
hence no classical probability theory can be applied there.

Let me first make some remarks on the question of whether an exotic
probability theory might explain away the Bell paradox.
Though there is no direct relation, I am reminded of an earlier attempt by 
\citet{pitowsky89} to resolve all paradoxes through adopting a mathematically
very sophisticated and non-standard version of probability theory, in that case,
by allowing non-measurable random variables and events. 
If events are not measurable, and moreover have lower and upper probabilities equal to
zero and one respectively, then relative frequencies do not converge, but
can have all values between $0$ and $1$ as points of accumulation. This allows
\citet{pitowsky89} to wriggle out of the constraint of Bell's inequality. Each probability
concerning hidden variables can take any value. 

Now experimentalists know that relative frequencies of macroscopic outcomes do
tend to converge under many repetitions of a carefully controlled experiment,
whether in quantum mechanics or not. The proof of Bell's theorem as I give it
does not require stabilization of relative frequencies of some further
unspecified micro-variables, but of joint relative frequencies of 
macroscopic variables, both ``what was actually measured'' and of
``what might have been measured''.  Moreover it assumes that the stabilized
values respect, by showing statistical independence, the physical independence
which follows from locality.  The results of a coin toss on one side of Innsbruck
campus is not correlated with a photon measurement on the other side.
In the case of  \citet{pitowsky89}, exotic probability does not  ``explain'' at all;
what is called an explanation is sleight-of-hand hidden under
impressive (but very specialistic) mathematics. 
At best, the explanation would imply a physics which is even more weird
than quantum mechanics.

I have yet to study the case for $p$-adic probability carefully, but a priori I am
highly sceptical.

Regarding Kolmogorov and von Mises I have already remarked that I do
not see any opposition between alternative views of probability here. Komogorov
merely describes probability, von Mises tries to explain it. Komogorov's theory is
mere accountancy. The underlying variable $\omega$ of a Komogorovian
probability space is not a physical cause, a hidden variable, it is merely a label
of a possible outcome. Naturally, in classical physical systems, there is a many-to-one
correspondence between initial conditions and distinguishable final conditions,
so one \emph{could} think of $\omega$ as being an element of a big list of initial
configurations. But this is not obligatory and, outside of physics, it is not usual.
See \citet{kolmogorov33} for very clear descriptions of what $\omega$ is
supposed to stand for and how probability can be interpreted. I think you will find
that Kolmogorov was definitely a contextualist.

\subsection*{{Kracklauer and the bombs under Bell's theory}}

According to Kracklauer, one counter-example is enough to explode a theorem.
Not content with one bomb he has come up with local realistic explanations of
a large number of celebrated experiments in quantum mechanics.
Unfortunately, showing that a long list of historical experiments did not \emph{prove}
what various experimenters and interpreters claim, does not prove a
certain theory, which inspired those experiments, wrong.

On the theoretical side he also has a large number of arguments,
but in my opinion none is persuasive.  One is that in real experiments there
are not binary outcomes but there is macroscopic photoelectric current. 
But one can convert a continuous current to a binary outcome (does it exceed a given
threshold or not). Bell's argument just requires that binary outcomes are output
and analysed; any intermediate steps are irrelevant.

Another argument is that photons do not actually exist. This certainly is a serious
point regarding Bell-type experiments in quantum optics, and is connected to
the Fifth Position,  to which I will return. 
As a mathematician I have to admit that the word ``photon" is perhaps no
more than just a word.
What we call a photon is associated with certain mathematical objects in certain
theories of   ``electro-magnetic radiation''
and associated with point-like events which one can identify
in various experiments involving ``light''.  
Mathematics itself is just a game of logical manipulations of distinct symbols on
pieces of paper.  Bell was careful to describe his decisive experiment in terms of
macroscopic every-day laboratory objects, and avoided any use of words like
``particle'' which only have a meaning within an existing theory.

Another argument is that the mathematics of spin does not involve Planck's
constant hence does not involve quantum mechanics. The transfer of EPR
to the realm of spin half or of photons is lethal. However, it seems to me that quantum
mechanics is as much about incompatible observables as about Planck's
constant.

Finally, Kracklauer enlists the support of the \citet{jaynes89}, who claimed to have
resolved all probability paradoxes in physics
by proper use of probability theory. According to E.T.~Jaynes,
Bell's factorization was an improper use of the chain rule for conditional probability.
Apparently Jaynes did not recognise an uncontroversial use of the notion of
conditional independence. Suppose I have a large collection of pairs of dice.
The two dice in each pair are identical. However half the pairs have two $1$'s, two $2$'s
and two $3$'s on their faces, the half have two $4$'s, two $5$'s
and two $6$'s. Call these Type 1 and Type 2 dice. Naturally if many times in
succession,
we take a random pair of dice,
send one to Amsterdam and the other to Bagdad, and toss each dice once,
there will be a strong correlation between the outcomes in the two locations.
Denote by $X$ and $Y$ the outcomes at the two locations, and by $T$ the
type of the dice. Suppose moreover that the dice-throwing apparatus in
Amsterdam and Bagdad each depend on a setting, called $a$ and $b$, 
which is chosen by a technician
in each laboratory. (The result of the setting is to bias the outcome
in a way which I will not further specify here.)
Bell calculates as follows:
\begin{align*}
\mathrm E_{ab}\{XY\}~&=~\mathrm E\{\ \mathrm E_{ab}\{ XY\mid T\}\,\}\\
                  &=~\Pr\{T=1\}\,\mathrm E_{ab}\{ XY\mid T=1\}~+~\Pr\{T=2\} \,\mathrm E_{ab}\{ XY\mid T=2\}\\
                  &=~\Pr\{T=1\}\,\mathrm E_a\{ X\mid T=1\}\,\mathrm E_b\{ Y\mid T=1\}\\
                  &~~~~~~~~~~~+~\Pr\{T=2\}\,\mathrm E_a\{ X\mid T=2\}\,\mathrm E_b\{ Y\mid T=2\}.
\end{align*}
Jaynes prefers to consider probabilities than expectations, that is fine.
He points out that the mere fact that our probability of seeing a particular value for $X$
is immediately changed when we are told the outcome of $Y$, does not mean
any spooky action at a distance (as Bell also many times explained).
He is also willing to apply the definition of conditional probability to write
\begin{align*} \Pr\nolimits_{ab}\{X=x&, Y=y\mid T=t\}\\
~&=~\Pr\nolimits_{ab}\{X=x\mid Y=y, T=t\}\,\Pr\nolimits_{ab}\{Y=y\mid T=t\}
\end{align*}
but then refuses to admit
\begin{align*}
\Pr\nolimits_{ab}\{X=x\mid Y=y,T=t\}~&=~\Pr\nolimits_a\{X=x\mid T=t\},\\
\Pr\nolimits_{ab}\{Y=y\mid T=t\}~&=~\Pr\nolimits_b\{Y=y\mid T=t\},
\end{align*}
going on to say that Bell's theorem only prohibits Bell's kind of local hidden variable
models, not all.  He does not make any attempt to specify what he understands 
by a local model, and expresses great surprise at very new
results of Steve Gull, presented at the same conference as Jaynes' own paper,
in which a computer network metaphor is introduced and where it is shown that
the singlet correlations cannot be simulated on such a network!
(Steve Gull faxed me his two pages of notes on this, which he likes to use an examination
exercise. His proof uses Fourier analysis). Jaynes thought
that it would take another $30$ years to understand Gull's work, just as
it had taken the world $20$ years to understand Bell's (the decisive understanding 
having just come from E.T.). I am not impressed. 

Bell's use of probability language was in 1964 still a bit clumsy. 
Jaynes'  work led him to a strong sense that any
probability paradox in physics is most likely the result of muddled thinking.
I suspect that Jaynes was so confident of this general rule
that he made no attempt 
to understand Bell's argument
and consequently completely missed the point.

\subsection*{{Volovich and the fifth position}}

Volovich's recent work shows that in an EPR type context of the state of
two entangled particles propagating in three-dimensional space, quantum
mechanics itself would prohibit a loophole free test of local realism.
Basically, particles will be lost with a too large probability, and the detection
loophole is present.

In my opinion it would be interesting to find out if this is generic. However one
must bear in mind that Bell's theorem is not dependent on a particular kind
of physical scenario (for instance, polarization of entangled photons). The
mathematical analysis must be carried out at a much more fundamental level in
order to show that no physical system consisting of two well separated
subsystems can evolve into a sufficiently entangled state by any means whatsover.

I would rather expect progress here to come from 't Hooft's programme: show that
quantum mechanics at the Planck scale has a local realistic explanation, show that
quantum mechanics at our scale is a consequence, and hence that it too is
constrained by local realism. 

Alternatively progress will come from experiment: someone does carry out a 
loophole free Bell-CHSH type experiment, or does factor large integers in
no time at all using a quantum computer.

\section*{{6. Last Word}}

Tossing a coin, shuffling a pack or cards, picking a ball from an urn, are classical
paradigms of randomness. Moreover all these experiments are well understood
both from a physical and from a mathematical point of view. We understand perfectly
well how small variations in initial conditions  are magnified exponentially and
result in quite unpredictable macroscopic results. On the basis of physical symmetries
we can propose uniform probability distributions over initial conditions, when
listed appropriately, and can use this to predict the probabilities of macroscopic
outcomes, for instance of biased roulette wheels. We understand moreover that
the probabilities of the macroscopic outcomes are remarkably robust to the
probability distribution of initial conditions. Finally, the probability conclusions are
quite independent of the flavour of probability interpretation.

Actually, generating a pseudo-random number on a computer is no different, except
that the fine control which we can impose on initial conditions and on each intermediate
step means that the result is exactly reproducible. But one can also buy a coin-tossing
apparatus which so precisely fixes the initial velocity and angular momentum (among other
factors) of the coin being tossed, that (unless one is unfortunate and chooses
initial conditions close to the boundary between ``heads'' and ``tails'' that the
coin falls the same way, (almost) every time.

That statistical independence holds when well separated physical systems are
each used to generate randomness, is not harder to understand. An extraordinarily
exquisite coordination between the number of times a pack of cards is shuffled,
and between the force used to spin a coin into the air, could produce any degree
of correlation in their outcomes.

These considerations mean that for me, that Bell's theorem has more or less nothing
to do with interpretations of probability. Classical physical randomness, and classical
physical independence, are what are at stake. My conclusion (excluding the
fifth position) is that quantum mechanics is definitely non-classical.

In order to establish that quantum mechanics is non classical, we had to assume
that physical independence between randomization devices at separate locations
in space is possible. We had to assume a degree of control on the amount of
information passing from one physical system to another: when we press the button
labelled ``$1$'' on one of the measurement devices, only the fact that it was that button
and not the other is important for the subsequent physics, even though actually we
exert more or less pressure, for a longer or shorter time, and thereby could unbeknown
to us be introducing information from other locations and from the distant past into
the apparatus. Bell's conditional independence assumption is a way to express
the physical intuition, that even though this might introduce more statistical variation
into the outcome, it cannot carry information from the other wing of the experiment,
concerning the randomization outcome there.

I find it fascinating that in order to prove that quantum mechanics is intrinsically
probabilistic (the outcomes cannot be traced back to variation in initial conditions)
we must assume that we can ourselves generate randomness. And in order to
demonstrate the kind of non-separatbility implied by entanglement, we have to assume
control and separation of the physical systems which we use in our experiments.

\section*{{References}}

\bibliographystyle{Chicago}

\renewcommand\bibsection{{}}


\section*{{Appendix 1: Weihs' data}}

\begin{center}
\begin{tabular}{|r | r || r | r || r | r ||}
\hline
 &  & $b=1$  & $b=1$  &  $b=2$  & $b=2$ \\
 \hline
 &  & $y=+1$ & $y=-1$ & $ y=+1$ & $y=-1$\\
 \hline
 \hline
 $a=1$ & $x=+1$ & $ 313 $&  $1728 $ & $1636 $&$179$\\
 \hline
  $a=1$ & $x=-1$ & $ 1978   $ & $ 351$ & $ 294 $ & $ 1143$ \\
 \hline
 \hline
$a=2$ & $x=+1$ & $  418 $ & $  1683 $ & $ 269  $ & $ 1100 $\\
\hline
$a=2$ & $x=-1$ & $1578  $ & $  361  $ & $ 1386  $ & $  156 $\\
\hline
\hline
\end{tabular}
\end{center}

\noindent The table show the numbers of occurrences of each
of the $16$ possible values of $(a,b,x,y)$, see Weihs' 1999 thesis,
page 113, available from his personal web pages 
at \texttt{www.quantum.at}.
The grand total is $N=14\,573$.

\section*{{Appendix 2: A local model of the singlet correlations}}

I present a caricature of the Hess-Philipp model, \texttt{quant-ph/0212085}.
The caricature has all those properties,
on the basis of which Hess and Philipp claimed its
locality. However, the caricature is blatantly non local. 
This makes it clear that Hess and Philipp are only checking necessary conditions,
not sufficient conditions, for locality.
In my construction I will only consider planar settings (orientiations),
and measure angles as fractions of $2\pi$, thus settings $a$, $b$
become points in the unit interval $[0,1]$ with endpoints identified.
I am going to construct random variables $R$, $\Lambda^*$, $\Lambda^{**}$,
$\Lambda$ whose joint probability distribution is allowed by Hess and Philipp to
depend on $a$ and $b$.  Actually, my $R$ will be a $2$-vector.
$R$ is supposed to be some kind of microscopic (i.e., hidden to the experimenter)
time variable. $\Lambda^*$ and $\Lambda^{**}$ are station variables. $\Lambda$ 
is a source variable, transmitted to both stations.

Let $a$ and $b$ be given.
Let $\Lambda^*$, $\Lambda^{**}$, and $\Lambda$
be independent random variables, each uniformly distributed on $[0,1]$. 
Define $R=(R_1,R_2)$
as follows:
\begin{equation}
R_1~=~(\Lambda^{**}+a)~\text{mod}~1,
\end{equation}
\begin{equation}
R_2~=~(\Lambda^{*}+b)~\text{mod}~1,
\end{equation}
As required by HP, conditional on
$R$, the pair $(\Lambda^*,\Lambda^{**})$ is independent of $\Lambda$.
All further independence properties desired by HP are 
trivially satisfied. However,
\begin{equation}
b~=~(R_2-\Lambda^*)~\text{mod}~1,
\end{equation}
\begin{equation}
a~=~(R_1-\Lambda^{**})~\text{mod}~1.
\end{equation}
Consequently, given $R$ and $\Lambda^*$ one can reconstruct $b$;
 given $R$ and $\Lambda^{**}$ one can reconstruct $a$ and $\Lambda^*$.
 
 Finally, let $A=A(\Lambda^*,\Lambda,R,a)$ and $B=B(\Lambda^*,\Lambda,R,b)$ 
 be functions
 taking values in $\{-1,+1\}$.
From the given arguments to $A$ and $B$, the missing station setting $b$ and $a$
can be reconstructed. From $a$, $b$ and $\Lambda$ one can construct a pair
of binary random variables with joint probability distribution depending in any way
one likes on $a$ and $b$. In particular one can arrange to
reproduce the singlet correlations.

To prove that both the HP model and this caricature are non-local, it suffices to
observe that they reproduce the singlet correlations in a realistic fashion,
and therefore by Bell's theorem cannot be local-realistic. However, according
to Hess and Philipp this conclusion is 
short-sighted. Obviously, $R$ is not an element of
reality! The only elements of reality in my model are $\Lambda$, $\Lambda^*$
and $\Lambda^{**}$.  They are evidently local, so my model is local, after all.

\end{document}